\documentclass[reprint,superscriptaddress,
amsmath,amssymb,aps,pra,twocolumn]{revtex4-2}
\usepackage{graphicx}
\usepackage{dcolumn}
\usepackage{newtxtext}
\usepackage{newtxmath}
\usepackage{amsmath,amssymb}
\usepackage{physics}
\usepackage{bm}     
\usepackage[breaklinks=true,colorlinks,citecolor=blue,linkcolor=blue,urlcolor=blue]{hyperref}
\usepackage{xcolor}
\usepackage{orcidlink}
\usepackage{natbib}

\newcommand{\ad}{\hat a^{\dagger}}
\newcommand{\bd}{\hat b^{\dagger}}
\newcommand{\ah}{\hat a}
\newcommand{\bh}{\hat b}
\newcommand{\ii}{\mathrm{i}}
\newcommand{\ee}{\mathrm{e}}
\newcommand{\gEP}{g_{\mathrm{EP}}}
\newcommand{\Heff}{\hat H_{\mathrm{eff}}}
\newcommand{\expec}[1]{\langle #1 \rangle}

\newif\ifshowrevisions
\showrevisionstrue

\ifshowrevisions
  \newenvironment{addendum}{\begingroup\color{red}}{\endgroup}
  \newenvironment{revision}{\begingroup\color{red}}{\endgroup}
  
\else

\fi

\begin{document}
\title{Unified Spectral, Dynamical, and Correlation Signatures of an Exceptional Point in Cavity Optomechanics}
\author{Khazali~\surname{Fahmi}\orcidlink{0000-0001-9943-3137}}\email[Corresponding author: ]{khazali.fahmi@ui.ac.id}
\affiliation{Department of Physics, Faculty of Mathematics and Natural Sciences, Universitas Indonesia, Depok 16424, Indonesia}
\affiliation{Research Center for Quantum Physics, National Research and Innovation Agency (BRIN), South Tangerang 15314, Indonesia}
\affiliation{Physics Study Program, Faculty of Military Mathematics and Natural Sciences, Republic of Indonesia Defense University, Bogor 16810, Indonesia}
\author{Ahmad R. T. \surname{Nugraha}\orcidlink{0000-0002-5108-1467}}\email{ahmad.ridwan.tresna.nugraha@brin.go.id}
\affiliation{Research Center for Quantum Physics, National Research and Innovation Agency (BRIN), South Tangerang 15314, Indonesia}
\affiliation{Department of Engineering Physics, School of Electrical Engineering, Telkom University, Bandung 40257, Indonesia}
\author{Ferry A. A. \surname{Nugroho}\orcidlink{0000-0001-5571-0454}}\email{ferryanggoroardynugroho@sci.ui.ac.id}
\affiliation{Department of Physics, Faculty of Mathematics and Natural Sciences, Universitas Indonesia, Depok 16424, Indonesia}
\author{Adam B. \surname{Cahaya}\orcidlink{0000-0002-2068-9613}}\email[Corresponding author: ]{adam@sci.ui.ac.id}
\affiliation{Department of Physics, Faculty of Mathematics and Natural Sciences, Universitas Indonesia, Depok 16424, Indonesia}
\begin{abstract}
We investigate how the exceptional-point structure of a dissipative cavity optomechanical system is inherited by experimentally accessible dynamical, spectral, and quantum-statistical observables. At optical-mechanical resonance, the effective non-Hermitian first-moment dynamics exhibits a second-order exceptional point, where two eigenvalues and their eigenvectors coalesce and the eigenvalue splitting follows the characteristic square-root dependence on perturbations. We show that the same square-root feature also governs transient photon and phonon populations, first-order coherence, spectral poles, and second-order intensity correlations. Below the exceptional point, the dynamics is non-oscillatory, whereas above it damped oscillations emerge together with frequency splitting of the spectral poles. At the exceptional point, the Jordan-block structure produces polynomial-exponential relaxation and a second-order spectral pole. Using the quantum regression theorem and Gaussian moment factorization, we further show that the stationary fluctuations satisfy the Siegert relation linking first- and second-order correlations. Although the zero-delay autocorrelations retain their thermal value, the finite-delay intensity correlations exhibit clear exceptional-point signatures. This finding connects non-Hermitian mode coalescence with measurable dynamical and correlation observables in cavity optomechanics.
\end{abstract}
\date{\today}
\maketitle
\section{Introduction}
In recent years, non-Hermitian physics has emerged as an active area of research, with exceptional points (EPs) representing one of its most intriguing phenomena. Exceptional points are non-Hermitian degeneracies at which two or more eigenvalues and their associated eigenvectors coalesce~\cite{kato1966perturbation,heiss2012physics,bender1998real,bender2024pt,downing2025first}. EPs have attracted significant interest in both fundamental studies and practical applications. They have been demonstrated in diverse platforms, including microcavities~\cite{dembowski2001experimental,dembowski2003observation,dembowski2004encircling}, coupled resonators~\cite{nikzamir2022highly,nada2017theory,zhang2018phonon}, superconducting circuits~\cite{partanen2019exceptional}, and cavity optomechanical systems~\cite{jing2017high,xiong2021higher,ozdemir2019parity,xiong2022higher}. These systems have opened new opportunities for applications such as loss-induced transparency~\cite{dong2025exceptional,zhang2018loss}, coherent perfect absorption~\cite{wang2021coherent,horner2024coherent}, and ultrasensitive sensing~\cite{wiersig2020review,chen2017exceptional}.

A key feature of exceptional points is their profound impact on the spectral response and dynamical behavior of non-Hermitian systems. Near an EP, the eigenvalue splitting induced by a small perturbation no longer follows the conventional linear dependence but instead exhibits a characteristic square-root scaling for second-order EPs, leading to a significantly enhanced spectral response to external perturbations~\cite{wiersig2016sensors}. Beyond that, exceptional points also modify the temporal evolution of dissipative systems. The coalescence of eigenvalues and eigenvectors gives rise to nonexponential dynamics associated with the Jordan-block structure at the EP, providing a dynamical signature that distinguishes the EP from the regimes away from it. Such dynamical features can manifest themselves in quantities connected to experimentally accessible observables, including photon and phonon populations, first-order coherence functions, and the corresponding optical and mechanical spectra~\cite{arkhipov2020liouvillian}.

Despite significant progress in understanding exceptional-point physics in dissipative and optomechanical systems, most previous studies have focused on the eigenmode structure, mean-field amplitudes, populations, stability, or $\mathcal{PT}$-symmetric dynamics~\cite{ryu2015exceptional,xu2021optomechanical}. A less explored yet important problem is how the exceptional-point structure of the underlying non-Hermitian modes is inherited by different levels of quantum observables, ranging from equal-time populations to two-time coherence functions, emission spectra, and higher-order intensity correlations. This distinction is crucial because those quantities probe different aspects of the open-system dynamics. The first moments determine the effective non-Hermitian mode structure, whereas higher-order moments and correlation functions characterize the distribution and temporal evolution of quantum fluctuations. Establishing a direct connection among these observables therefore provides a more complete and experimentally relevant characterization of exceptional-point dynamics.

In this work, we investigate a unified connection between the exceptional point of the effective non-Hermitian first-moment dynamics and experimentally accessible quantum observables in a dissipative cavity-optomechanical system. We first identify the second-order exceptional point of the effective Hamiltonian and characterize its square-root eigenvalue response. We then show how the same non-Hermitian structure is inherited by the steady-state and transient photon and phonon populations, first-order coherence functions, optical and mechanical spectra, and second-order intensity correlations.  We find that the dynamical behavior across these different levels of quantum observables can be determined by a single parameter that combines optomechanical coupling strength, optical decay rate, and mechanical damping rate.

\section{Model and Equations of Motion}

\begin{figure}[tb]
    \centering
    \includegraphics[width=1.0\columnwidth]{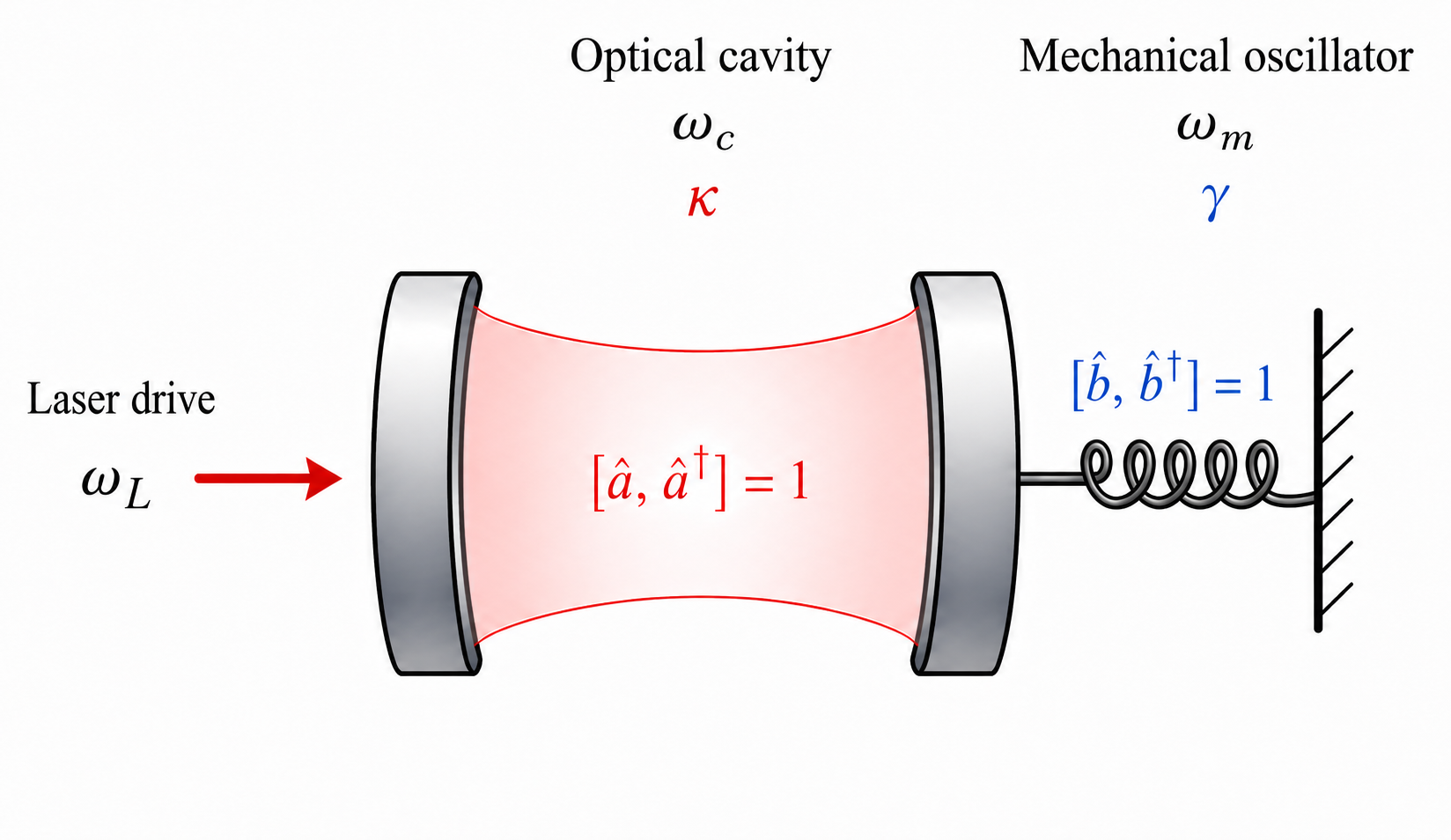}
    \caption{Schematic of the Fabry-P\'erot cavity optomechanical system. An optical cavity with resonance frequency $\omega_c$ and decay rate $\kappa$ is driven by a laser at frequency $\omega_L$ and coupled, with strength $g$, to a mechanical oscillator of frequency $\omega_m$ and damping rate $\gamma$.}
    \label{fig:optomechanics}
\end{figure}

We consider a dissipative cavity optomechanical system consisting of a single optical mode coupled to a single
mechanical oscillator, as sketched in Fig.~\ref{fig:optomechanics}. The optical cavity mode is described by the annihilation (creation) operator $\hat{a} (\hat{a}^\dagger)$, while the mechanical mode is described
by $\hat{b} (\hat{b}^\dagger)$. These operators satisfy $\comm{\hat{a}}{\hat{a}^\dagger} = 1$, and $\comm{\hat{b}}{\hat{b}^\dagger} = 1$. The optical and mechanical resonance frequencies are denoted by $\omega_c$ and $\omega_m$, respectively. In the frame rotating at the drive frequency $\omega_L$, the Hamiltonian reads ($\hbar = 1$)
\begin{align}
    \hat{H} &= \Delta_{\rm c} \hat{a}^\dagger \hat{a} +\omega_m \hat{b}^\dagger \hat{b} + g_0 \hat{a}^\dag \hat{a} (\hat{b} + \hat{b}^\dag) + \mathrm{i} \Omega_{\rm L} (\hat{a}^\dagger - \hat{a})\label{eq:optomechanics-Hamiltonian},
\end{align}
where $\Delta_{\rm c} = \omega_c - \omega_{\rm L}$ is the cavity–laser detuning, $g_0$ is the single-photon optomechanical coupling strength, and $\Omega_{\rm L}$ is the drive amplitude.

Under a strong coherent drive, the optical and mechanical operators may be displaced into their steady-state amplitudes and quantum fluctuations,
\begin{align}
    \hat a &= \alpha_{\rm ss} + \delta \hat a, \qquad \hat b = \beta_{\rm ss} + \delta \hat b,
    \label{eq:displace}
\end{align}
where $\alpha_{\rm ss}=\langle \hat a\rangle_{\rm ss}$ and $\beta_{\rm ss}=\langle \hat b\rangle_{\rm ss}$ denote the steady-state coherent amplitudes of the optical and mechanical modes, respectively, while $\delta\hat a$ and $\delta\hat b$ describe quantum fluctuations around these mean values. The corresponding steady-state amplitudes are
\begin{align}
    \alpha_{\rm ss} &= \frac{\Omega_{\mathrm{L}}}{\kappa/2 + \ii \Delta}, \qquad \beta_{\mathrm{ss}} = \frac{-\mathrm{i} g_0 \abs{\alpha_{\mathrm{ss}}}^2}{\gamma/2 + \ii \omega_m},
    \label{eq:meanfield}
\end{align}
where $\Delta = \Delta_{\rm c} + 2g_0\,\mathrm{Re}\,\beta_{\rm ss}$ is the effective cavity detuning. Neglecting the nonlinear fluctuation terms and applying the rotating-wave approximation appropriate for a red-detuned drive ($\Delta \simeq \omega_m > 0$) yields the linearized beam-splitter Hamiltonian
\begin{align}
    \hat{H} &\simeq \Delta \hat{a}^\dagger \hat{a} + \omega_m \hat{b}^\dagger \hat{b} + g (\hat{a}^\dagger \hat{b} + \hat{a} \hat{b}^\dagger), \label{eq:OP-4}
\end{align}
where $g=g_0\alpha_{\rm ss}$ is the field-enhanced (linearized) optomechanical coupling. The drive phase is chosen so that $g$ is real. In what follows we drop the symbol $\delta$ and denote the fluctuation operators simply by $\hat a$ and $\hat b$.

The open-system dynamics of density matrix $\hat{\rho}$ obey the Lindblad master equation~\cite{campaioli2024quantum}
\begin{align}
    \dot{\hat\rho} &= -\mathrm{i}[\hat H,\hat \rho] +\frac{\kappa}{2}\,\mathcal{D}[\hat a]\hat \rho +\frac{\gamma}{2}(n_{\rm th}+1)\,\mathcal{D}[\hat b]\hat \rho \nonumber\\
    &\quad +\frac{\gamma}{2}\,n_{\rm th}\,\mathcal{D}[\hat b^\dagger]\hat \rho,
    \label{eq:master}
\end{align}
with the dissipator $\mathcal{D}[\hat o]\hat\rho = 2\hat o \hat\rho \hat o^{\dagger} - \hat o^{\dagger} \hat o \hat\rho - \hat\rho \hat o^{\dagger} \hat o$, optical decay rate $\kappa$, mechanical damping rate $\gamma$, and mean thermal occupation $n_{\rm th}$ of the mechanical bath (the optical bath is taken at zero temperature). Equation~\eqref{eq:master} is the starting point for both the first- and second-order dynamical equations studied below.

Using $\dfrac{d}{dt} \langle \hat O \rangle = \mathrm{Tr} \{\hat O\dot{\hat\rho}\}$, the first moments obey
\begin{align}
    \frac{d}{dt} \langle \hat{a} \rangle &= \left(-\mathrm{i} \Delta - \frac{\kappa}{2}\right) \langle \hat{a} \rangle - \mathrm{i} g \langle \hat{b} \rangle \label{eq:field_a-4}\\
    \frac{d}{dt} \langle \hat{b} \rangle &= -\mathrm{i} g \langle \hat{a} \rangle + \left(-\mathrm{i} \omega_m - \frac{\gamma}{2}\right) \langle \hat{b} \rangle. 
\end{align}
These equations can be written as a Schr\"odinger-like equation
\begin{align}
    \mathrm{i} \frac{d\psi}{dt} &= \hat{H}_{\rm eff} \psi, \qquad \psi = \big(\langle \hat a \rangle, \langle \hat b \rangle\big)^{\rm T},
    \label{eq:schrolike}
\end{align}
with the effective non-Hermitian Hamiltonian
\begin{align}
    \hat{H}_{\rm eff} &= 
    \begin{pmatrix}
        \Delta - \mathrm{i} \frac{\kappa}{2} & g \\
        g & \omega_m - \mathrm{i} \frac{\gamma}{2}
    \end{pmatrix}.
    \label{eq:heff}
\end{align}
The non-Hermiticity of $\hat H_{\mathrm{eff}}$, arising from optical and mechanical dissipation, renders its eigenvalues complex. Their real and imaginary parts encode the oscillation frequencies and decay rates of the hybridized optomechanical modes and form the basis for the emergence of exceptional points.  Note that equations~\eqref{eq:master}--\eqref{eq:heff} are utilized to obtain all the results below, both analytically and numerically.   The Lindblad master equation~\eqref{eq:master} in particular is solved using truncated Fock bases for the optical and mechanical modes~\cite{johansson2012qutip}. 

\section{Results and Discussion}

\subsection{Eigenvalue Splitting near the Exceptional Point}
\label{sec:eigs}

Exceptional points are characterized not only by the coalescence of eigenvalues and eigenvectors but also by their extraordinary response to external perturbations. In contrast to ordinary Hermitian degeneracies, where the eigenvalue splitting varies linearly with the perturbation strength, non-Hermitian exceptional points exhibit a square-root dependence. This non-analytic behavior underlies the enhanced spectral sensitivity that constitutes one of the most important signatures of exceptional-point physics.

For the resonant condition $\Delta = \omega_{m} \equiv \omega$, the eigenvalues of $\hat H_{\mathrm{eff}}$ are
\begin{align}
    \lambda_{\pm} &=\omega -\ii\,\frac{\kappa+\gamma}{4}\pm\sqrt{g^{2} - \gEP^2},
    \label{eq:lambdas}
\end{align}
where the exceptional-point coupling is
\begin{align}
    \gEP &=\frac{|\kappa-\gamma|}{4}.
    \label{eq:gEP}
\end{align}
At $g = \gEP$ the two eigenvalues coalesce to $\lambda_0 = \omega - \ii (\kappa + \gamma) / 4$ and the corresponding eigenvectors become parallel, signalling a second-order exceptional point. The eigenvalue splitting is
\begin{align}
    \Delta\lambda = \lambda_+ - \lambda_- = 2\sqrt{g^{2} - \gEP^2}.
    \label{eq:split}
\end{align}

\begin{figure}[tb]
\centering
\includegraphics[width=\columnwidth]{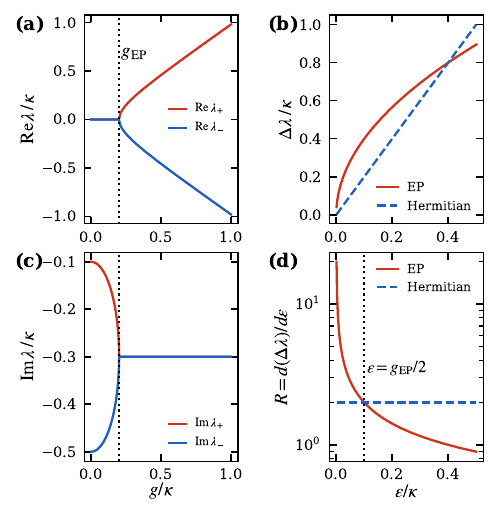}
\caption{(a) Real and (c) imaginary parts of the eigenvalues $\lambda_\pm$ of $\Heff$ as a function of the coupling strength $g$ for the resonant condition $\Delta=\omega_m$. The dotted vertical line marks the exceptional point $\gEP=|\kappa-\gamma|/4$. Below the EP the imaginary parts split while the real parts remain degenerate and above the EP the roles are exchanged. Parameters used are $\gamma=0.2\kappa$ and the common resonant frequency $\omega = \Delta = \omega_m$ is taken as the zero-frequency reference. (b) Eigenvalue splitting versus perturbation $\epsilon$. The EP follows the square-root law $\sqrt{8\gEP\epsilon}$ whereas the Hermitian degeneracy responds linearly, $2\epsilon$. (d) Responsivity $R=d(\Delta\lambda)/d\epsilon$ on a logarithmic scale. The EP response diverges as $\epsilon\to0$ and crosses the constant Hermitian value $R'=2$ at $\epsilon=\gEP/2$.}
\label{fig:eval-sensi}
\end{figure}

As shown in Figs.~\ref{fig:eval-sensi}(a) and~\ref{fig:eval-sensi}(c), the exceptional point separates two qualitatively different spectral regimes. For $g<\gEP$, the square root is imaginary, so the real parts of $\lambda_\pm$ remain degenerate while their imaginary parts split.
The two hybrid modes therefore oscillate at a common frequency but decay at different rates. For $g>\gEP$, the square root is real, so the imaginary parts become degenerate while the real parts split symmetrically. The two modes therefore have distinct frequencies but share the same decay rate. The exceptional point at $g = \gEP$ therefore marks the onset of real-frequency splitting of the complex poles. A visibly resolved spectral doublet, however, generally emerges only sufficiently above the EP~\cite{aspelmeyer2014cavity}.

To quantify the spectral response near the exceptional point, we introduce a small perturbation $\epsilon$ to the coupling strength, $g=\gEP+\epsilon$ with $\epsilon\ll\gEP$. Substituting into Eq.~\eqref{eq:split} gives, to leading order,
\begin{equation}
\Delta\lambda\simeq\sqrt{8\gEP\,\epsilon},
\label{eq:sqrtresp}
\end{equation}
so the eigenvalues near the EP become $\lambda_\pm=\lambda_0\pm\sqrt{2\gEP\epsilon}$. The square-root dependence implies that the eigenvalue splitting decreases more slowly with decreasing perturbation strength than in the corresponding Hermitian case. Consequently, its derivative with respect to the perturbation becomes singular as $\epsilon\to0$.

For comparison, consider the Hermitian limit obtained by neglecting dissipation ($\kappa=\gamma=0$) at resonance, for which
\begin{align}
    \Heff &= 
    \begin{pmatrix}
        \omega & g\\
        g & \omega
    \end{pmatrix}
\end{align}
with eigenvalues $\lambda'_\pm=\omega\pm g$ and splitting $(\Delta\lambda)'=2g$. Introducing the same perturbation $g=\epsilon$ gives $(\Delta\lambda)'=2\epsilon$, which grows linearly and remains finite for arbitrarily small perturbations. The enhanced eigenvalue response can be quantified through the spectral responsivities
\begin{equation}
R=\frac{d(\Delta\lambda)}{d\epsilon}=\sqrt{\frac{2\gEP}{\epsilon}},\qquad
R'=\frac{d(\Delta\lambda)'}{d\epsilon}=2 .
\label{eq:responsivity}
\end{equation}
As $\epsilon\to0$, the exceptional-point responsivity $R$ diverges, whereas the Hermitian responsivity $R'$ remains constant. The two responses coincide at $\epsilon=\gEP/2$, below which the exceptional-point system exhibits a larger eigenvalue responsivity [see Fig.~\ref{fig:eval-sensi}(b) and Fig.~\ref{fig:eval-sensi}(d)]. This singular behavior originates from the square-root topology of the exceptional point and demonstrates the enhanced eigenvalue response characteristic of EP-based sensing schemes.

\subsection{Steady-State Population and Intermode Coherence}
\label{subsec:steady}

\begin{figure}[tb]
    \centering
    \includegraphics[width=\columnwidth]{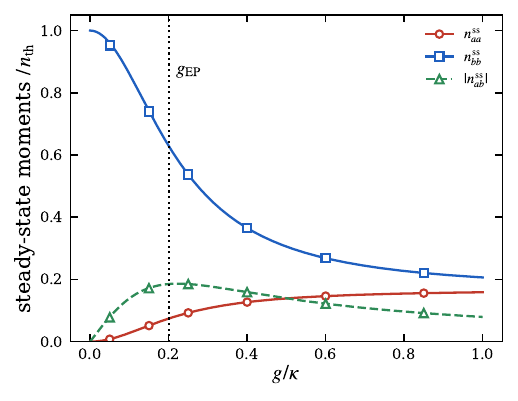}
    \caption{Steady-state photon population $n^{\mathrm{ss}}_{aa}$, phonon population $n^{\mathrm{ss}}_{bb}$, and intermode coherence $|n^{\mathrm{ss}}_{ab}|$ (all in units of $n_{\mathrm{th}}$) versus coupling strength $g$ at resonance ($\delta=0$). Solid curves are the analytic results, Eqs.~\eqref{eq:naa}--\eqref{eq:nab}. Open symbols are the numerical steady state of the master equation. The dotted line marks $\gEP$. Parameters selected are as those in Fig.~\ref{fig:eval-sensi}.}
    \label{fig:ss}
\end{figure}

While the eigenvalue analysis reveals the spectral structure of $\Heff$, the experimentally measurable quantities are the expectation values of physical observables. The relevant second-order moments are the photon and phonon populations, $\expec{\ad\ah}$ and $\expec{\bd\bh})$, as well as the coherence established between the optical and mechanical modes, that is $\expec{\ad\bh}$ and $\expec{\bd\ah}$. Collecting them in the vector
\begin{equation}
\mathbf{V}=\big(\expec{\ad\ah},\;\expec{\bd\bh},\;\expec{\ad\bh},\;\expec{\bd\ah}\big)^{\mathrm{T}},
\end{equation}
the master equation~\eqref{eq:master} yields a closed linear system
\begin{equation}
\dot{\mathbf V}=\mathbf D\,\mathbf V+\mathbf F,
\label{eq:Vdot}
\end{equation}
where the drift matrix,
\begin{align}
    \mathbf D &= 
    \begin{pmatrix}
        -\kappa & 0 & -\ii g & \ii g\\
        0 & -\gamma & \ii g & -\ii g\\
        -\ii g & \ii g & \ii\delta -\Gamma & 0\\
        \ii g & -\ii g & 0 & -\ii\delta -\Gamma
    \end{pmatrix},
\end{align}
describes the coupled dissipative dynamics and the inhomogeneous term $\mathbf F = (0,\gamma n_{\mathrm{th}},0,0)^{\mathrm{T}}$ accounts for thermal excitation of the mechanical reservoir. Imposing $\dot{\mathbf V}=0$ and solving the resulting algebraic system gives the steady-state populations $n^{\mathrm{ss}}_{aa}=\expec{\ad\ah}_{\mathrm{ss}}$, $n^{\mathrm{ss}}_{bb}=\expec{\bd\bh}_{\mathrm{ss}}$ and coherences $n^{\mathrm{ss}}_{ab}=\expec{\ad\bh}_{\mathrm{ss}}$, $n^{\mathrm{ss}}_{ba}=\expec{\bd\ah}_{\mathrm{ss}}=(n^{\mathrm{ss}}_{ab})^{*}$,
\begin{align}
    n^{\mathrm{ss}}_{aa}&=\frac{g^{2}\gamma(\kappa+\gamma)}{(\Gamma^{2}+\delta^{2})\kappa\gamma+g^{2}(\kappa+\gamma)^{2}}\,n_{\mathrm{th}},
    \label{eq:naa}\\[4pt]
    n^{\mathrm{ss}}_{bb}&=\frac{(\Gamma^{2}+\delta^{2})\kappa\gamma+g^{2}(\kappa+\gamma)\gamma}{(\Gamma^{2}+\delta^{2})\kappa\gamma+g^{2}(\kappa+\gamma)^{2}}\,n_{\mathrm{th}},
    \label{eq:nbb}\\[4pt]
    n^{\mathrm{ss}}_{ab}&=\frac{\ii g\kappa\gamma(\Gamma^{2}+\delta^{2})}{(\Gamma-\ii\delta)\big[(\Gamma^{2}+\delta^{2})\kappa\gamma+g^{2}(\kappa+\gamma)^{2}\big]}\,n_{\mathrm{th}},
    \label{eq:nab}\\[4pt]
    n^{\mathrm{ss}}_{ba}&=\frac{-\ii g\kappa\gamma(\Gamma^{2}+\delta^{2})}{(\Gamma+\ii\delta)\big[(\Gamma^{2}+\delta^{2})\kappa\gamma+g^{2}(\kappa+\gamma)^{2}\big]}\,n_{\mathrm{th}},
    \label{eq:nba}
\end{align}
where
\begin{equation}
\Gamma=\frac{\kappa+\gamma}{2},\qquad \delta=\Delta-\omega_{m} .
\end{equation}
These expressions describe how the thermal excitation of the mechanical reservoir is redistributed between the two modes by the optomechanical interaction. Here, $n^{\mathrm{ss}}_{aa}$ and $n^{\mathrm{ss}}_{bb}$ are the stationary photon and phonon populations, respectively, while $n^{\mathrm{ss}}_{ab}$ quantifies the steady-state intermode coherence. In the uncoupled limit $g\to0$, we recover $n^{\mathrm{ss}}_{bb}\to n_{\mathrm{th}}$ and $n^{\mathrm{ss}}_{aa}\to0$, as expected.  With the increase in coupling strength, the mechanical population decreases monotonically, whereas the optical population increases and approaches a finite saturation value. In contrast, the intermode coherence $|n^{\mathrm{ss}}_{ab}|$ exhibits a nonmonotonic dependence on $g$ and reaches its maximum at $g=\sqrt{\kappa\gamma}/2$. For the parameters used in Fig.~\ref{fig:ss}, this maximum occurs at $g \simeq 0.224\kappa$, slightly above the exceptional point. The analytical curves are indistinguishable from a direct numerical solution of the steady state of Eq.~\eqref{eq:master}.

An important observation is that none of these equal-time steady-state quantities exhibits a singular feature exactly at $g = \gEP$. The photon and phonon populations vary smoothly across the exceptional point, while the intermode coherence reaches its maximum at a coupling that does not generally coincide with $\gEP$. The exceptional point is therefore not necessarily revealed by a sharp anomaly in stationary equal-time observables.

\subsection{Transient Population Dynamics}
\label{sec:transient}

The steady-state solutions describe the long-time behavior but not how equilibrium is approached. To study the relaxation, we introduce the deviation from the steady state, $\tilde{\mathbf V}=\mathbf V(t)-\mathbf V_{\mathrm{ss}}$, where $t$ denotes the evolution time measured from the choosen initial condition at $t=0$. The deviation obeys the homogeneous equation $\dot{\tilde{\mathbf V}}=\mathbf D\tilde{\mathbf V}$. At resonance ($\delta=0$) the components $\tilde n_{ij}(t)=n_{ij}(t)-n^{\mathrm{ss}}_{ij}$ satisfy
\begin{equation}
\frac{d}{dt}
    \begin{pmatrix}\tilde n_{aa}\\ \tilde n_{bb}\\ \tilde n_{ab}\\ \tilde n_{ba}\end{pmatrix} =
    \begin{pmatrix}
        -\kappa & 0 & -\ii g & \ii g\\
        0 & -\gamma & \ii g & -\ii g\\
        -\ii g & \ii g & -\Gamma & 0\\
        \ii g & -\ii g & 0 & -\Gamma
    \end{pmatrix}
    \begin{pmatrix}
    \tilde n_{aa}\\ 
    \tilde n_{bb}\\ 
    \tilde n_{ab}\\ 
    \tilde n_{ba}
    \end{pmatrix}.
    \label{eq:transmat}
\end{equation}
Additional insight is gained by introducing the collective variables
\begin{align}
P&=\tilde n_{aa}+\tilde n_{bb}, & X&=\tilde n_{aa}-\tilde n_{bb},\\
Y&=\tilde n_{ab}-\tilde n_{ba}, & Z&=\tilde n_{ab}+\tilde n_{ba},
\end{align}
which represent the total population deviation, the population imbalance, and the antisymmetric and symmetric coherence combinations, respectively. In these variables Eq.~\eqref{eq:transmat} becomes
\begin{align}
\dot P&=-\Gamma P-\tfrac{d}{2}X,\label{eq:Pdot}\\
\dot X&=-\Gamma X-\tfrac{d}{2}P-2\ii g\,Y,\\
\dot Y&=-\Gamma Y-2\ii g\,X,\\
\dot Z&=-\Gamma Z,\label{eq:Zdot}
\end{align}
where $d=\kappa-\gamma$ denotes the difference between the optical decay and mechanical damping rates. The symmetric coherence $Z$ decouples and relaxes independently at the rate $\Gamma$, while $P$, $X$, and $Y$ remain coupled. Writing $\mathbf u=(P,X,Y,Z)^{\mathrm{T}}$, the system reads $\dot{\mathbf u}=(-\Gamma\mathbf{I}_4+\mathbf M)\mathbf u$, where $\mathbf{I}_4$ is the $4 \times 4$ identity matrix and
\begin{equation}
    \mathbf M =
    \begin{pmatrix}
        0 & -\tfrac{d}{2} & 0 & 0\\
        -\tfrac{d}{2} & 0 & -2\ii g & 0\\
        0 & -2\ii g & 0 & 0\\
        0 & 0 & 0 & 0
    \end{pmatrix}.
\end{equation}
The eigenvalues of the second-order-moment generator $-\Gamma\mathbf{I}_4+\mathbf{M}$ are
\begin{equation}
    \lambda_{\rm mom} = -\Gamma,
    \quad 
    -\Gamma,\quad
    -\Gamma+\frac{\Omega}{2},
    \quad
    -\Gamma-\frac{\Omega}{2},
    \label{eq:moment_eigenvalues}
\end{equation}
with
\begin{equation}
    \Omega=\sqrt{d^{2}-16g^{2}}.
    \label{eq:Omega}
\end{equation}
Thus, the same discriminant $\Omega$ that identifies the exceptional point in the first-moment dynamics also determines the relaxation spectrum of the second-order moments. The matrix $\mathbf M$ obeys the relation $\mathbf M^{3} =(\Omega^{2}/4)\,\mathbf M$, which allows the matrix exponential to be evaluated in closed form,
\begin{align}
    \mathbf u(t) &= \ee^{-\Gamma t}\!\left[\mathbf{I}_4+\frac{2\sinh(\Omega t/2)}{\Omega}\mathbf M \right. \nonumber\\
    &\quad  \left. + \frac{4\big(\cosh(\Omega t/2)-1\big)}{\Omega^{2}}\mathbf M^{2}\right]\!\mathbf u(0).
    \label{eq:usol}
\end{align}
The original moments follow from $n_{aa}=n^{\mathrm{ss}}_{aa}+(P+X)/2$, $n_{bb}=n^{\mathrm{ss}}_{bb}+(P-X)/2$, $n_{ab}=n^{\mathrm{ss}}_{ab}+(Z+Y)/2$, and $n_{ba}=n^{\mathrm{ss}}_{ba}+(Z-Y)/2$.

The same discriminant $\Omega$ that determines the exceptional-point condition also governs the transient second-order-moment dynamics. Using $g_{\rm EP}=|\kappa-\gamma|/4$, one has $\Omega=4\sqrt{g_{\rm EP}^2-g^2}$, so that $\Omega=0$ corresponds
precisely to the exceptional point $g=g_{\rm EP}$ identified in Sec.~\ref{sec:eigs}.

Three dynamical regimes follow directly from Eq.~\eqref{eq:usol}. For $g<g_{\rm EP}$, $\Omega$ is real and the transient dynamics is non-oscillatory, involving exponential contributions with decay rates
$\Gamma$, $\Gamma-\frac{\Omega}{2}$, and $\Gamma+\frac{\Omega}{2}$.
As $g$ approaches $g_{\rm EP}$ from below, the two distinct decay rates $\Gamma\mp\Omega/2$ coalesce at $\Gamma$.
For $g>g_{\rm EP}$, writing $\Omega=\ii\chi$ with
\begin{equation}
    \chi = \sqrt{16g^2-(\kappa-\gamma)^2} = 4\sqrt{g^2-g_{\rm EP}^2},
\end{equation}
the hyperbolic functions in Eq.~\eqref{eq:usol} become trigonometric. The second-order moments therefore exhibit damped oscillations with the characteristic population-oscillation frequency
\begin{equation}
    \omega_{\rm pop} = \frac{\chi}{2} = 2\sqrt{g^2-g_{\rm EP}^2}.
\end{equation}
At $g=g_{\rm EP}$, $\Omega\rightarrow0$, and the propagator reduces continuously to the polynomial form associated with the Jordan-block structure
\begin{equation}
    \mathbf{u}_{\rm EP}(t) = e^{-\Gamma t} \left[\mathbf{I}_4 +t\mathbf{M} +\frac{t^2}{2}\mathbf{M}^2\right] \mathbf{u}(0),
    \label{eq:u_EP}
\end{equation}
so that the exponential decay is multiplied by a polynomial in time. For the second-order-moment dynamics considered here, the Jordan structure generates terms proportional to $e^{-\Gamma t}$, $te^{-\Gamma t}$, and $t^2 e^{-\Gamma t}$. This polynomially modified relaxation constitutes the temporal signature of mode coalescence at the exceptional point. 

\begin{figure}[tb]
\centering
\includegraphics[width=\columnwidth]{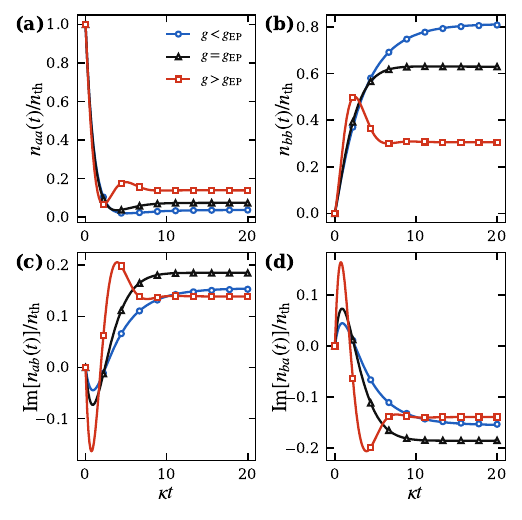}
\caption{Transient dynamics of the second-order moments in the three dynamical regimes. Panels (a) and (b) show the photon population $n_{aa}(t)$ and phonon population $n_{bb}(t)$, respectively, while panels (c) and (d) show the imaginary parts of the intermode coherences $n_{ab}(t)$ and $n_{ba}(t)$, respectively. The three regimes correspond to $g < \gEP$ (blue), $g = \gEP$ (black), and $g > \gEP$ (red), with $g=0.12\kappa$, $g=\gEP=0.2\kappa$, and $g=0.5\kappa$, respectively. The fluctuation modes are initially prepared in the state
$|1,0\rangle$, corresponding to one optical excitation and the mechanical ground state. Solid curves denote the analytical solutions of Eqs.~\eqref{eq:usol} and~\eqref{eq:u_EP}, while open symbols show the corresponding numerical results obtained from the Lindblad master equation. Parameters selected are as those in Fig.~\ref{fig:eval-sensi}.}
\label{fig:transient}
\end{figure}

Figure~\ref{fig:transient} depicts how the exceptional-point structure is inherited by the transient second-order moments. In Fig.~\ref{fig:transient}(a), the photon population $n_{aa}(t)$ starts from unity because the fluctuation modes are initially prepared in the state $|1,0\rangle$. For $g < \gEP$, the optical excitation rapidly decays and subsequently approaches its steady-state value without sustained oscillations. At $g = \gEP$, the relaxation remains non-oscillatory but is governed by the polynomial-exponential form of Eq.~\eqref{eq:u_EP}, reflecting the coalescence of the transient decay modes. For $g > \gEP$, the photon population exhibits damped oscillatory behavior. This behavior reflects the exchange of excitation between the optical and mechanical modes, while dissipation progressively suppresses the oscillation amplitude.

Figure~\ref{fig:transient}(b) shows the evolution of the phonon population $n_{bb}(t)$. Starting from the mechanical ground state, the phonon population increases as the initial optical excitation is transferred to the mechanical mode. For $g < \gEP$, the phonon population relaxes monotonically toward the steady state without oscillating. At $g = \gEP$, the separate decay modes coalesce into a critically damped relaxation governed by polynomial-exponential decay. For $g > \gEP$, $n_{bb}$ develops an overshoot before settling to its steady-state value, providing a direct population signature of the oscillatory regime. The long-time values in Figs.~\ref{fig:transient}(a) and~\ref{fig:transient}(b) differ because the steady-state photon and phonon populations vary with the coupling strength, according to Eqs.~\eqref{eq:naa} and~\eqref{eq:nbb}.

The intermode coherence is depicted in Figs.~\ref{fig:transient}(c) and~\ref{fig:transient}(d). The imaginary part of $n_{ab}(t)$, shown in Fig.~\ref{fig:transient}(c), is initially zero because the two modes are uncorrelated in the initial state. The interaction dynamically generates optical-mechanical coherence, producing an initial transient variation followed by relaxation toward the nonzero steady-state coherence. The transient becomes increasingly pronounced as the coupling is increased, and for $g > \gEP$ the sign change and subsequent overshoot reflect the coherent exchange associated with the split hybrid modes. Figure~\ref{fig:transient}(d) displays imaginary part of $n_{ba}(t)$. Since $n_{ba}(t) = n_{ab}^*(t)$, its imaginary part satisfies $\Im[n_{ba}(t)] = -\Im[n_{ab}(t)]$, explaining the mirror-like behavior of Figs.~\ref{fig:transient}(c) and~\ref{fig:transient}(d).

Taken together, the EP therefore marks the critical boundary between non-oscillatory and oscillatory relaxation, where the distinct transient decay modes coalesce. Importantly, the $t^2e^{-\Gamma t}$ term in Eq.~\eqref{eq:u_EP} does not imply a third-order EP of the original effective Hamiltonian, but arises from the Jordan structure of the enlarged second order-moment dynamics.


\subsection{First-order coherence}
\label{sec:g1}

The second-order moments discussed in the previous section describe the population dynamics and intermode coherence at equal times, but they do not fully characterize the temporal coherence of the optical and mechanical fields. The latter is quantified by the normalized first-order correlation function
\begin{equation}
    g^{(1)}_{AB}(\tau) = \lim_{t\rightarrow\infty} \frac{\left\langle \hat{A}^{\dagger}(t)\hat{B}(t+\tau) \right\rangle}{\sqrt{ \left\langle \hat{A}^{\dagger}\hat{A}\right\rangle_{\rm ss} \left\langle \hat{B}^{\dagger}\hat{B}\right\rangle_{\rm ss}}},
    \label{eq:g1def}
\end{equation}
where $A,B\in\{a,b\}$. Here, $t$ denotes the reference observation time, while $\tau$ is the time delay separating the two field operation. The limit $t \rightarrow \infty$ ensures that the system has reached its stationary state, so that the two-time correlation become independent of the absolute reference time $t$ and depends only on the relative delay $\tau$. This definition includes the optical and mechanical autocorrelations, $g^{(1)}_{aa}(\tau)$ and
$g^{(1)}_{bb}(\tau)$, as well as the optical and mechanical cross correlations $g^{(1)}_{ab}(\tau)$ and $g^{(1)}_{ba}(\tau)$.

It is convenient to collect the stationary two-time correlations into
the matrix
\begin{equation}
    \mathbf{G}(\tau) =
    \begin{pmatrix}
        \left\langle \hat{a}^{\dagger}(0)\hat{a}(\tau) \right\rangle_{\rm ss} & \left\langle \hat{b}^{\dagger}(0)\hat{a}(\tau) \right\rangle_{\rm ss}\\[2mm]
        \left\langle \hat{a}^{\dagger}(0)\hat{b}(\tau) \right\rangle_{\rm ss} & \left\langle \hat{b}^{\dagger}(0)\hat{b}(\tau) \right\rangle_{\rm ss}
    \end{pmatrix}.
    \label{eq:Gmatrix}
\end{equation}
At zero delay this matrix is determined entirely by the steady-state second moments derived in Sec.~\ref{subsec:steady},
\begin{equation}
    \mathbf{G}(0) =
    \begin{pmatrix}
        n^{\rm ss}_{aa} & n^{\rm ss}_{ba}\\
        n^{\rm ss}_{ab} & n^{\rm ss}_{bb}
    \end{pmatrix}.
    \label{eq:G0}
\end{equation}
The individual normalized correlation functions therefore follow from
\begin{align}
    g^{(1)}_{aa}(\tau) &= \frac{G_{11}(\tau)}{n^{\rm ss}_{aa}},
    & g^{(1)}_{bb}(\tau) &= \frac{G_{22}(\tau)}{n^{\rm ss}_{bb}},
    \label{eq:g1_auto}\\
    g^{(1)}_{ab}(\tau) &= \frac{G_{21}(\tau)}{\sqrt{n^{\rm ss}_{aa}n^{\rm ss}_{bb}}}, & g^{(1)}_{ba}(\tau) &= \frac{G_{12}(\tau)} {\sqrt{n^{\rm ss}_{aa}n^{\rm ss}_{bb}}}.
    \label{eq:g1_cross}
\end{align}

By the quantum regression theorem, the two-time correlations obey the
same linear equations as the first moments. Consequently,
\begin{equation}
    \frac{\partial}{\partial\tau}\mathbf{G}(\tau) = \mathbf{A}\mathbf{G}(\tau),
    \label{eq:G_eom}
\end{equation}
where
\begin{equation}
    \mathbf{A} =
    \begin{pmatrix}
        -\ii \Delta - \kappa/2 & - \ii g\\
        -\ii g & - \ii \omega_m - \gamma/2
    \end{pmatrix}.
    \label{eq:A_matrix}
\end{equation}
At optical-mechanical resonance, $\Delta=\omega_m\equiv\omega$, the dynamical matrix can be decomposed as
\begin{equation}
    \mathbf{A} = -\left(\frac{\Gamma}{2} + \ii \omega\right)\mathbf{I}_{2} + \mathbf{N},
    \label{eq:A_decomp}
\end{equation}
where $\mathbf{I}_2$ is the $2 \times 2$ identity matrix and
\begin{equation}
    \mathbf{N} =
    \begin{pmatrix}
        -d/4 & -\ii g\\
        -\ii g & d/4
    \end{pmatrix}.
    \label{eq:N_matrix}
\end{equation}
The matrix $\mathbf{N}$ obeys relation $\mathbf{N}^{2} = \frac{\Omega^{2}}{16}\mathbf{I}_{2}$, which allows the matrix exponential to be evaluated in closed form,
\begin{align}
    \mathbf{G}(\tau) &= e^{-(\Gamma/2 + \ii\omega)\tau} \left[\cosh\left(\frac{\Omega\tau}{4}\right)\mathbf{I}_{2} \right. \nonumber\\
    &\quad \left. + \frac{4}{\Omega} \sinh\left(\frac{\Omega\tau}{4}\right)\mathbf{N} \right] \mathbf{G}(0).
    \label{eq:G_general}
\end{align}
Thus, the same discriminant $\Omega$ that governs the transient population dynamics also controls the first-order coherence. Equation~(\ref{eq:G_general}) immediately reveals three qualitatively different regimes separated by the exceptional point.

For $g<g_{\rm EP}$, the discriminant is real and positive. Writing
\begin{equation}
    q\equiv\Omega = \sqrt{d^{2}-16g^{2}},
    \label{eq:q_below}
\end{equation}
Equation~(\ref{eq:G_general}) can be written as
\begin{align}
    \mathbf{G}(\tau) = \frac{e^{-\ii\omega\tau}}{2} \Bigg[&\left(\mathbf{I}_{2} + \frac{4\mathbf{N}}{q}\right) e^{-\left(\Gamma/2-q/4\right)\tau} \nonumber\\
    + & \left(\mathbf{I}_{2} - \frac{4\mathbf{N}}{q}\right) e^{-\left(\Gamma/2+q/4\right)\tau}\Bigg] \mathbf{G}(0).
    \label{eq:G_below_EP}
\end{align}
The correlation functions therefore contain two purely decaying dynamical contributions with decay rates
\begin{equation}
    \Gamma_{\pm}' = \frac{\Gamma}{2} \pm \frac{q}{4}.
    \label{eq:g1_decay_below}
\end{equation}
Both contributions oscillate only with the common carrier frequency $\omega$, and no splitting of the oscillation frequencies occurs. Consequently, the coherence dynamics is non-oscillatory after removal of the common carrier phase. The cross-correlations, however, may still display nonmonotonic features because the two decay channels enter with different complex amplitudes.

At $g=g_{\rm EP}$, $\Omega \rightarrow 0$, the exponential in
Eq.~(\ref{eq:G_general}) then reduces to
\begin{equation}
    \mathbf{G}_{\rm EP}(\tau) = e^{-(\Gamma/2+\ii\omega)\tau} \left( \mathbf{I}_{2} + \tau\mathbf{N} \right) \mathbf{G}(0).
    \label{eq:G_EP}
\end{equation}
Hence, every first-order correlation function has the generic form
\begin{equation}
    g^{(1)}_{AB,{\rm EP}}(\tau) = \left( C_{AB} + D_{AB}\tau\right) e^{-(\Gamma/2+\ii\omega)\tau},
    \label{eq:g1_generic_EP}
\end{equation}
where $C_{AB}$ and $D_{AB}$ are determined by the corresponding steady-state correlations. The term linear in $\tau$ is a direct signature of the Jordan-block structure associated with eigenvector coalescence. In contrast to an ordinary exponential relaxation, the EP therefore produces a critically damped polynomial exponential coherence dynamics.

For $g>g_{\rm EP}$, $\Omega$ becomes purely imaginary. We write
\begin{equation}
    \Omega=\ii\chi,
    \qquad
    \chi = \sqrt{16g^{2}-d^{2}} = 4\sqrt{g^{2}-g_{\rm EP}^{2}}.
    \label{eq:chi}
\end{equation}
Equation~(\ref{eq:G_general}) becomes
\begin{align}
    \mathbf{G}(\tau) &= e^{-(\Gamma/2+\ii\omega)\tau} \left[\cos\left(\frac{\chi\tau}{4}\right)\mathbf{I}_{2} \right. \nonumber\\
    & \quad \left. + \frac{4}{\chi} \sin\left(\frac{\chi\tau}{4}\right)\mathbf{N}\right] \mathbf{G}(0).
    \label{eq:G_above_EP}
\end{align}
Thus, the correlation amplitudes acquire damped oscillatory contributions with the characteristic coherence-oscillation frequency
\begin{equation}
    \omega_{\rm coh} = \frac{\chi}{4} = \sqrt{g^{2}-g_{\rm EP}^{2}}.
    \label{eq:omega_coh}
\end{equation}
Equivalently, the time-domain correlations contain two frequency
components centered at
\begin{equation}
    \omega_{\pm} = \omega \pm \frac{\chi}{4} = \omega \pm \sqrt{g^{2}-g_{\rm EP}^{2}},
    \label{eq:omega_pm}
\end{equation}
whose separation is
\begin{equation}
    \omega_{+}-\omega_{-} = \frac{\chi}{2} = 2\sqrt{g^{2}-g_{\rm EP}^{2}}.
    \label{eq:freq_sep}
\end{equation}
This separation is identical to the real-part eigenvalue splitting above the exceptional point.

Figure~\ref{fig:g1} shows how the same three dynamical regimes appear in the normalized first-order coherence functions. Figure~\ref{fig:g1}(a) presents the optical autocorrelation $\abs{g_{aa}^{(1)}(\tau)}$ and starts from unity at $\tau = 0$. For $g < \gEP$, the correlation decays smoothly as a superposition of two non-oscillatory contributions with different decay rates. At $g = \gEP$, these two rates coalesce and the correlation acquires the polynomial-exponential form of Eq.~\eqref{eq:g1_generic_EP}. Although the magnitude still appears as a smooth decay, its functional form is no longer a simple exponential. For $g > \gEP$, the correlation develops minima and partial revivals, which are the time-domain manifestation of interference between the two split-frequency components $\omega_\pm$.

The mechanical autocorrelation $\abs{g_{bb}^{(1)}(\tau)}$ in Fig~\ref{fig:g1}(b) exhibits the same transition. For $g < \gEP$, it decreases without oscillatory modulation, while at $g = \gEP$ its decay is governed by the coalesced mode and the linear polynomial prefactor. For $g > \gEP$, pronounced minima and revivals emerge. These oscillatory features occur with the characteristic coherence frequency $\omega_{\mathrm{coh}}$, demonstrating that the mechanical coherence inherits the same non-Hermitian mode splitting as the optical field.

Figures~\ref{fig:g1}(c) and~\ref{fig:g1}(d) display the cross-correlations $\abs{g_{ab}^{(1)}(\tau)}$ and $\abs{g_{ba}^{(1)}(\tau)}$, respectively. Unlike the autocorrelations, their zero-delay values are not constrained to unity, because they quantify the normalized coherence between two different modes. In Fig.~\ref{fig:g1}(c), the optical-to-mechanical correlation decreases smoothly for $g < \gEP$ and at $g = \gEP$, whereas For $g > \gEP$ it develops zeros and subsequent revivals as the two split-frequency contributions interfere destructively and constructively. Figure~\ref{fig:g1}(d) shows a related but temporally asymmetric behavior, even for $g < \gEP$ and at $g = \gEP$ a short-delay maximum can occur before the eventual decay. This does not represent an additional oscillatory mode; rather, it results from the interference of the two non-oscillatory decay channels with different amplitudes in the directional cross-correlation. For $g > \gEP$, clear minima and revivals again appear.

The difference between Figs.~\ref{fig:g1}(c) and~\ref{fig:g1}(d) emphasizes that the two cross-correlations probe different time orderings of optical-to-mechanical and mechanical-to-optical coherence transfer. Nevertheless, all four correlation functions are governed by the same discriminant $\Omega$. Thus, Fig.~\ref{fig:g1} provides a direct time-domain connection between the exceptional point of the first-moment effective Hamiltonian and experimentally relevant two-time coherence functions.


\subsection{Spectral signatures of the exceptional point}
\label{sec:spec}

The frequency-domain response follows directly from the stationary two-time correlation functions. We first define the complex one-sided Fourier--Laplace transform of the correlation matrix,
\begin{equation}
    \boldsymbol{\mathcal{S}}(\nu) = \int_{0}^{\infty} d\tau\, 
    e^{\ii\nu\tau} \mathbf{G}(\tau),
    \label{eq:S_complex}
\end{equation}
where $\nu$ denotes the spectral analysis frequency associated with time delay $\tau$ through Fourier transform. For the diagonal elements, the conventional stationary power spectrum
is
\begin{equation}
    S_{AA}(\nu) = 2\,{\rm Re}\,
    \mathcal{S}_{AA}(\nu),
    \qquad
    A\in\{a,b\},
    \label{eq:S_auto}
\end{equation}
whereas for the off-diagonal correlations we consider the real cross-spectral response
\begin{equation}
    S^{(R)}_{AB}(\nu) = 2\,{\rm Re}\, \mathcal{S}_{AB}(\nu),
    \qquad
    A\neq B.
    \label{eq:S_cross}
\end{equation}
The latter can contain both absorptive and dispersive contributions and is therefore not restricted to positive values.

To obtain the analytic spectrum, we define
\begin{equation}
    z(\nu) = \frac{\Gamma}{2} + \ii(\omega-\nu).
    \label{eq:z}
\end{equation}
For $\Omega\neq0$, substituting Eq.~(\ref{eq:G_general}) into Eq.~(\ref{eq:S_complex}) gives
\begin{equation}
    \boldsymbol{\mathcal{S}}(\nu) = \frac{1}{2} \left[\frac{\mathbf{I}_{2} + 4\mathbf{N}/\Omega}{z(\nu)-\Omega/4} + \frac{\mathbf{I}_{2} - 4\mathbf{N}/\Omega}{z(\nu)+\Omega/4}\right] \mathbf{G}(0).
    \label{eq:S_general}
\end{equation}
The poles of Eq.~(\ref{eq:S_general}) provide a direct frequency-domain representation of the same three dynamical regimes identified above.

For $g<g_{\rm EP}$, $\Omega=q$ is real, and the two poles are
\begin{equation}
    z(\nu) \mp \frac{q}{4} = \left(\frac{\Gamma}{2} \mp \frac{q}{4}\right) + \ii(\omega-\nu).
    \label{eq:poles_below}
\end{equation}
Both poles therefore have the same resonance frequency,
\begin{equation}
    \nu_1 = \nu_2=\omega,
\end{equation}
but different decay rates. Accordingly, the optical and mechanical spectra remain unsplit below the EP, with their line shapes determined by the superposition of two dissipative contributions.

Exactly at the exceptional point, the limit $\Omega\rightarrow0$ must be taken directly from the Jordan form in Eq.~(\ref{eq:G_EP}). The spectral matrix becomes
\begin{equation}
    \boldsymbol{\mathcal{S}}_{\rm EP}(\nu) =
    \left[\frac{\mathbf{I}_{2}}{z(\nu)} +
    \frac{\mathbf{N}}{z^{2}(\nu)} \right]
    \mathbf{G}(0).
    \label{eq:S_EP}
\end{equation}
The spectrum therefore contains both first- and second-order pole contributions. The term proportional to $z^{-2}$ is the frequency-domain counterpart of the linear-in-$\tau$ contribution in the first-order coherence. Thus, the spectrum remains unsplit at the EP, while the coalescence of the two dynamical modes produces a second-order spectral pole.

For $g>g_{\rm EP}$, writing $\Omega=\ii\chi$, the two poles acquire the same decay rate but split in resonance frequency,
\begin{equation}
    \nu_{\pm} = \omega \pm \frac{\chi}{4} = \omega \pm \sqrt{g^{2}-g_{\rm EP}^{2}}.
    \label{eq:spectral_centers_above}
\end{equation}
The corresponding frequency separation is
\begin{equation}
    \Delta\nu = \nu_{+}-\nu_{-} = \frac{\chi}{2} = 2\sqrt{g^{2}-g_{\rm EP}^{2}},
    \label{eq:spectral_splitting}
\end{equation}
which coincides with the real-part eigenvalue splitting obtained in Sec.~\ref{sec:eigs}. The exceptional point therefore marks the onset of real-frequency pole splitting.

\begin{figure}[tb]
    \centering
    \includegraphics[width=\columnwidth]{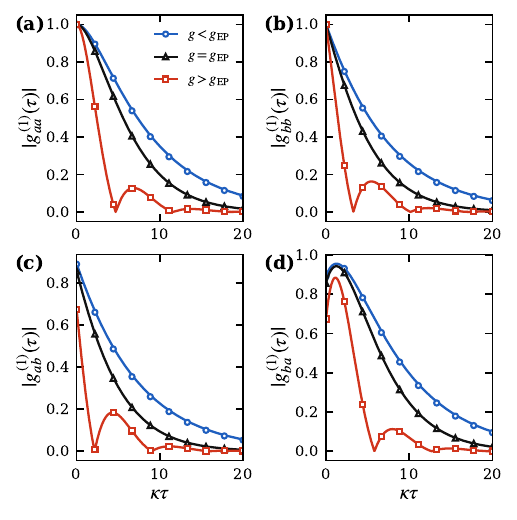}
    \caption{Magnitude of the normalized first-order coherence functions in the three dynamical regimes. Panels (a) and (b) show the optical and mechanical autocorrelations, $|g^{(1)}_{aa}(\tau)|$ and $|g^{(1)}_{bb}(\tau)|$, respectively, while panels (c) and (d) show the optical-to-mechanical and mechanical-to-optical cross-correlations, $|g^{(1)}_{ab}(\tau)|$ and $|g^{(1)}_{ba}(\tau)|$. The three regimes correspond to $g<g_{\rm EP}$, $g=g_{\rm EP}$, and $g>g_{\rm EP}$. Below the EP the correlations are governed by two non-oscillatory decay channels, at the EP they acquire a polynomial exponential form, and above the EP they exhibit damped oscillations. Solid curves represent the analytical results, while open symbols denote the numerical solutions of the Lindblad master equation. Parameters selected are as those in Fig.~\ref{fig:eval-sensi}.}
    \label{fig:g1}
\end{figure}

\begin{figure}[tb]
    \centering
    \includegraphics[width=\columnwidth]{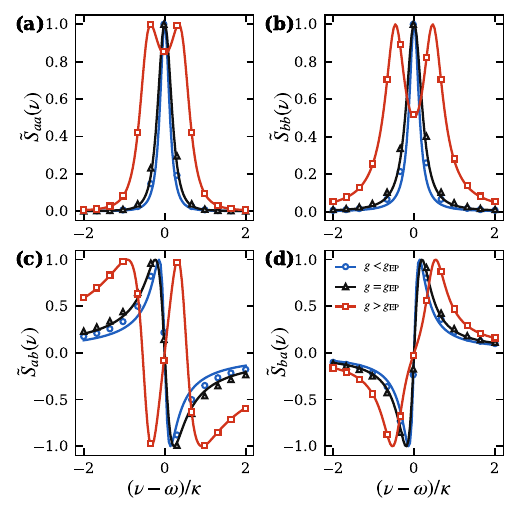}
    \caption{Normalized spectral responses corresponding to the first-order correlation functions in the three dynamical regimes. Panels (a) and (b) show the normalized optical and mechanical spectra, $\widetilde{S}_{aa}(\nu)$ and $\widetilde{S}_{bb}(\nu)$, respectively, while panels (c) and (d) show the normalized real cross-spectral responses, $\widetilde{S}_{ab}(\nu)$ and $\widetilde{S}_{ba}(\nu)$. Below the EP the two spectral poles share a common resonance frequency but have different decay rates. At the EP they coalesce into a second-order pole, whereas above the EP they split in resonance frequency while sharing the same decay rate. Each curve is independently normalized by its maximum absolute value. Solid curves represent the analytical results, while open symbols denote the corresponding numerical results. Parameters selected are as those in Fig.~\ref{fig:eval-sensi}.}
    \label{fig:spectrum}
\end{figure}

Figure~\ref{fig:spectrum} provides the frequency-domain counterpart of the coherence dynamics shown in Fig.~\ref{fig:g1}. Figure~\ref{fig:spectrum}(a) shows the normalized optical spectrum $\widetilde{S}_{aa}(\nu)$. For $g < \gEP$, the two dynamical poles have the same resonance frequency and therefore produce a single unsplit spectral feature centered at $\nu = \omega$, although the two contributions possess different decay rates. At $g = \gEP$, the poles coalesce and the spectrum remains centered at the same frequency. Its line shape, however, is modified by the second-order-pole contribution $\mathbf{N}/z^2(\nu)$ in Eq.~\eqref{eq:S_EP}. For $g > \gEP$, two spectral maxima emerge on opposite sides of the central frequency, reflecting the real-frequency splitting of the hybridized optomechanical modes.

Figure~\ref{fig:spectrum}(b) shows the corresponding normalized mechanical spectrum $\widetilde{S}_{bb}(\nu)$. The same unsplit-to-split transition is observed: a single resonance occurs below and at the EP, whereas a clear doublet develops above the EP. The two maxima are associated with the frequencies $\nu_\pm$ and their separation therefore follows $\Delta\nu$, exactly reproducing the real-part eigenvalue splitting of the effective non-Hermitian Hamiltonian. The optical and mechanical spectra thus provide two complementary experimental signatures of the same exceptional-point transition.

The real optical-to-mechanical cross-spectral response $\widetilde{S}_{ab}(\nu)$ is shown in Fig.~\ref{fig:spectrum}(c). In contrast to the diagonal spectra, the cross spectrum is not required to be positive because it contains both absorptive and dispersive contributions. Consequently, positive and negative spectral lobes occur around the resonance. For $g < \gEP$, these structures originate from two poles located at the same resonance frequency but having different decay rates. At $g = \gEP$, their coalescence and the resulting second-order pole produce a sharper variation of the cross-spectral response around the central resonance. For $g > \gEP$, the frequency splitting reorganizes the positive and negative extrema around the two hybrid-mode frequencies, producing the multi-lobed structure visible in the red curve.

Figure~\ref{fig:spectrum}(d) presents the reciprocal mechanical-to-optical cross-spectral response $\widetilde{S}_{ba}(\nu)$. It displays a complementary dispersive line shape with negative and positive contributions on opposite sides of the central resonance. As in Fig.~\ref{fig:spectrum}(c), the response remains associated with a common resonance frequency for $g < \gEP$, acquires the coalesced-pole structure at $g = \gEP$, and develops features associated with two separated resonances for $g > \gEP$. The differences between Figs.~\ref{fig:spectrum}(c) and~\ref{fig:spectrum}(d) arise from the directional ordering of the optical-mechanical two-time correlations and from the unequal optical and mechanical damping rates.

Therefore, Figs.~\ref{fig:spectrum}(a) -~\ref{fig:spectrum}(d) demonstrate that the frequency-domain observables retain the same exceptional-point topology found in the eigenvalue and time-domain analyses: decay-rate splitting below the EP, pole coalescence at the EP, and real-frequency splitting above it. Because every curve in Fig.~\ref{fig:spectrum} is independently normalized to its own maximum absolute value, the figure should be interpreted as a comparison of spectral line shapes, pole positions, and splitting, rather than of absolute spectral intensities.


\subsection{Second-order correlations and intensity fluctuations}
\label{sec:g2}

While the first-order correlation functions characterize the temporal coherence and spectral response of the optical and mechanical fields, the statistics of their intensity fluctuations are described by the
normalized second-order correlation function
\begin{equation}
    g^{(2)}_{AB}(\tau) = \lim_{t\rightarrow\infty} \frac{\left\langle \hat A^\dagger(t) \hat B^\dagger(t+\tau) \hat B(t+\tau) \hat A(t) \right\rangle}{\left\langle\hat A^\dagger\hat A\right\rangle_{\rm ss} \left\langle\hat B^\dagger\hat B\right\rangle_{\rm ss}},
    \label{eq:g2_def}
\end{equation}
where $A,B\in\{a,b\}$. The autocorrelations $g^{(2)}_{aa}(\tau)$ and
$g^{(2)}_{bb}(\tau)$ describe the photon and phonon intensity statistics, respectively, while $g^{(2)}_{ab}(\tau)$ and $g^{(2)}_{ba}(\tau)$ quantify the time-ordered intensity correlations between the optical and mechanical modes.

Because the coherent drive has been displaced into the classical steady-state amplitude, the fluctuation operators $\hat a$ and $\hat b$ have vanishing mean values. Furthermore, the linearized beam-splitter Hamiltonian conserves the total excitation number and does not generate anomalous correlations from the thermal mechanical bath and the vacuum optical bath. The stationary state of the fluctuations is therefore a zero-mean Gaussian state whose statistical properties are fully determined by the second moments.

For a zero-mean Gaussian state, Wick's theorem factorizes the four-operator correlation function according to
\begin{align}
    & \left\langle \hat A^\dagger(0) \hat B^\dagger(\tau) \hat B(\tau) \hat A(0) \right\rangle_{\rm ss} \nonumber\\
    &\qquad = n_A^{\rm ss} n_B^{\rm ss} + \left| \left\langle \hat A^\dagger(0)\hat B(\tau) \right\rangle_{\rm ss} \right|^2 + \left| \left\langle \hat A^\dagger(0)\hat B^\dagger(\tau) \right\rangle_{\rm ss} \right|^2,
    \label{eq:wick_general}
\end{align}
where
\begin{equation}
    n_a^{\rm ss}=n_{aa}^{\rm ss},
    \qquad
    n_b^{\rm ss}=n_{bb}^{\rm ss}.
\end{equation}
For the passive beam-splitter interaction considered here, the anomalous correlations vanish,
\begin{equation}
    \left\langle
    \hat A^\dagger(0)\hat B^\dagger(\tau)
    \right\rangle_{\rm ss}=0,
    \label{eq:anomalous_zero}
\end{equation}
so that Eq.~(\ref{eq:wick_general}) reduces to
\begin{equation}
    \left\langle \hat A^\dagger(0) \hat B^\dagger(\tau) \hat B(\tau) \hat A(0) \right\rangle_{\rm ss} = n_A^{\rm ss}n_B^{\rm ss} + \left| G_{AB}(\tau) \right|^2,
    \label{eq:wick_reduced}
\end{equation}
with
\begin{equation}
    G_{AB}(\tau) = \left\langle \hat A^\dagger(0)\hat B(\tau) \right\rangle_{\rm ss}.
    \label{eq:GAB}
\end{equation}

Using the normalized first-order coherence function defined in Eq.~(\ref{eq:g1def}), one obtains the Siegert relation
\begin{equation}
    g^{(2)}_{AB}(\tau) = 1 + \left| g^{(1)}_{AB}(\tau) \right|^2.
    \label{eq:siegert}
\end{equation}
Thus, in the present linear Gaussian system, the second-order correlation functions contain no independent dynamical scale beyond that already present in the first order coherence. In particular, the same exceptional-point discriminant
\begin{equation}
    \Omega = \sqrt{(\kappa-\gamma)^2-16g^2}
    \label{eq:Omega_g2}
\end{equation}
governs both quantities. Nevertheless, because $g^{(2)}-1$ depends on the modulus squared of $g^{(1)}$, the exceptional-point structure appears differently in the intensity fluctuations.

For the autocorrelations, the normalization implies
\begin{equation}
    g^{(1)}_{aa}(0) = g^{(1)}_{bb}(0) = 1,
\end{equation}
and hence
\begin{equation}
    g^{(2)}_{aa}(0) = g^{(2)}_{bb}(0) = 2.
    \label{eq:g2_auto_zero}
\end{equation}
The zero-delay photon and phonon statistics therefore remain thermal and bunched for all coupling strengths. The proximity to the exceptional point does not change the instantaneous autocorrelation from its thermal value. For the cross-correlations, in contrast,
\begin{align}
    g^{(2)}_{ab}(0) &= 1 + \frac{|n_{ab}^{\rm ss}|^2}{n_{aa}^{\rm ss}n_{bb}^{\rm ss}},\\
    g^{(2)}_{ba}(0) &= 1 + \frac{|n_{ba}^{\rm ss}|^2}{n_{aa}^{\rm ss}n_{bb}^{\rm ss}}.
    \label{eq:g2_cross_zero}
\end{align}
Since $n_{ba}^{\rm ss}=(n_{ab}^{\rm ss})^*$, the two cross-correlations have the same value at zero delay. Moreover, substituting the steady-state moments in Eqs.~\eqref{eq:naa} --~\eqref{eq:nba} and imposing the resonant condition $\delta = 0$ yields
\begin{equation}
    g^{(2)}_{ab}(0) = g^{(2)}_{ba}(0) = 1 + \frac{\kappa^2}{\kappa (\kappa + \gamma) + 4g^2}.
    \label{eq:g2_cross_resonance}
\end{equation} 
Thus, unlike the autocorrelations, whose zero-delay value remains fixed at 2, the equal-time cross-correlation depends continuously on the optomechanical coupling $g$. This dependence is smooth across $\gEP$ and therefore does not by itself constitute a singular signature of the exceptional point. The EP signature instead appears in the finite-delay temporal dynamics.

Through the Siegert relation, the three dynamical regimes obtained for
$g^{(1)}_{AB}(\tau)$ in Sec.~\ref{sec:g1} are directly inherited by $g^{(2)}_{AB}(\tau)$. Below the EP, the intensity correlations are governed by non-oscillatory combinations of exponential decay channels. At the EP, the first-order coherence has the generic Jordan form
\begin{equation}
    g^{(1)}_{AB,{\rm EP}}(\tau) = \left(A_{AB}+B_{AB}\tau\right) e^{-(\Gamma/2+\ii\omega)\tau},
\end{equation}
and hence
\begin{equation}
    g^{(2)}_{AB,{\rm EP}}(\tau) = 1 + \left( C_{0,AB} + C_{1,AB}\tau + C_{2,AB}\tau^2 \right)e^{-\Gamma\tau},
    \label{eq:g2_EP}
\end{equation}
where $C_{0,AB} = |A_{AB}|^2$, $C_{1,AB}=2\,{\rm Re}\!\left(A_{AB}^{*}B_{AB}\right)$, and $C_{2,AB} = |B_{AB}|^2$.
Thus, the linear Jordan prefactor of the first-order coherence becomes a quadratic polynomial exponential contribution in the intensity correlations.

Above the EP, writing $\Omega=\ii\chi$ with $\chi=4\sqrt{g^2-g_{\rm EP}^2}$, the modulus square in Eq.~(\ref{eq:siegert}) produces damped oscillations at the characteristic intensity-correlation frequency
\begin{equation}
    \omega_{\mathrm{corr}} = \frac{\chi}{2} = 2\sqrt{g^2-g_{\rm EP}^2}.
    \label{eq:g2_frequency}
\end{equation}
The same frequency also governs the oscillatory second-order-moment dynamics discussed in Sec.~\ref{sec:transient}.

\begin{figure}[tb]
\centering
\includegraphics[width=\columnwidth]{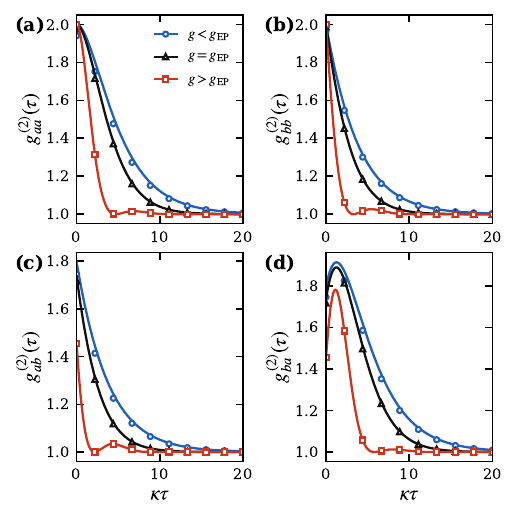}
\caption{Second-order correlation functions in the three dynamical regimes. Panels (a) and (b) show the optical and mechanical autocorrelations, $g_{aa}^{(2)}(\tau)$ and $g_{bb}^{(2)}(\tau)$, respectively, while panels (c) and (d) show the optical-to-mechanical and mechanical-to-optical cross-correlations, $g_{ab}^{(2)}(\tau)$ and $g_{ba}^{(2)}(\tau)$, respectively. The three regimes correspond to $g < \gEP$ (blue), $g = \gEP$ (black), and $g > \gEP$ (red). Solid curves represent the analytical results, while open symbols denote the corresponding numerical results. Parameters selected are as those in Fig.~\ref{fig:eval-sensi}.}
\label{fig:g2}
\end{figure}

Figure~\ref{fig:g2} summarizes how the exceptional-point dynamics is transferred to the intensity fluctuations. The optical and mechanical autocorrelations in Figs.~~\ref{fig:g2}(a) and~\ref{fig:g2}(b) start from the universal thermal value, $g^{(2)}_{aa}(0) = g^{(2)}_{bb}(0) = 2$ and approach the uncorrelated limit at long delay. Consequently, the exceptional point is not identified through the instantaneous autocorrelation, but through the temporal profile of its decay. Below the EP, the correlations relax through non-oscillatory decay channels associated with the distinct relaxation rates. At the EP, these rates coalesce and the Jordan structure produces the characteristic form. Above the EP, the correlation develops a damped oscillatory modulation with frequency in Eq.~\ref{eq:g2_frequency}. Because of the Siegert relation, these oscillations appears as modulations of the correlation excess above the asymptotic value rather than as oscillations about $g^{(2)} = 1$.

In Figs.~\ref{fig:g2}(c) and~\ref{fig:g2}(d), the two cross correlations start from the same zero-delay value given by Eq.~\eqref{eq:g2_cross_resonance}, which depends smoothly on $g$. They generally separate at finite delay because $g^{(2)}_{ab}(\tau)$ and $g^{(2)}_{ba}(\tau)$ correspond to opposite optical-to-mechanical and mechanical-to-optical temporal orderings. Despite this temporal-ordering asymmetry, both correlations exhibit the same EP-controlled transition from non-oscillatory relaxation below the EP to polynomial-exponential dynamics at the EP and damped oscillatory modulations above it.

\subsection{Experimental feasibility}

The exceptional point considered here is accessible within parameters already demonstrated in Fabry-P\'erot cavity optomechanics. The movable mechanical element in this platform consists of a low-stress silicon-nitride ($\mathrm{Si_3N_4}$) micromechanical resonator carrying a high-reflectivity dielectric Bragg mirror formed by alternating $\mathrm{Ta_2O_5}$ and $\mathrm{SiO_2}$ layers, which serves as one end mirror of the Fabry-P\'erot cavity~\cite{groblacher2009demonstration}. In the experiment of Gr\"oblacher \textit{et al.}~\cite{groblacher2009observation}, the reported parameters are a mechanical frequency $\omega_m/2\pi = 947~\mathrm{kHz}$, a cavity amplitude-decay rate $\kappa_G/2\pi = 215~\mathrm{kHz}$, corresponding to a cavity linewidth of $2\kappa_G/2\pi = 430~\mathrm{kHz}$, a mechanical linewidth $\gamma/2\pi = 0.14~\mathrm{kHz}$, and optomechanical couplings $g_G/2\pi = 78, 192, 260$, and $325~\mathrm{kHz}$ at drive powers of $0.6, 3.8, 6.9$, and $10.7~\mathrm{mW}$, respectively. Because the notation used in Ref.~\cite{groblacher2009observation} differs from that adopted here, these parameters are converted according to $\kappa = 2\kappa_G$ and $g = g_G/2$.

Using the converted parameters, Eq.~\eqref{eq:gEP} gives an exceptional-point coupling of $g_{\mathrm{EP}}/2\pi = 107.5~\mathrm{kHz}$. The experimental coupling at a drive power of $3.8~\mathrm{mW}$ corresponds to $g/2\pi=96~\mathrm{kHz}<g_{\mathrm{EP}}/2\pi$, whereas that at $6.9~\mathrm{mW}$ corresponds to $g/2\pi=130~\mathrm{kHz}>g_{\mathrm{EP}}/2\pi$, placing these two experimentally demonstrated operating points on opposite sides of the predicted EP. Within the near-resonant approximation $\Delta\simeq\omega_m$, the present theory therefore predicts no real-frequency pole splitting for the former case, while for $g/2\pi=130~\mathrm{kHz}$ the spectral poles yield a normal-mode separation. Using Eqs.~\eqref{eq:spectral_centers_above} and~\eqref{eq:spectral_splitting}, the normal-mode separation occurs at $\Delta\nu/2\pi = 146~\mathrm{kHz}$, corresponding to mode frequencies near $0.874$ and $1.020~\mathrm{MHz}$. At the largest reported coupling, $g/2\pi=162.5~\mathrm{kHz}$, the predicted separation increases to approximately $244~\mathrm{kHz}$, with mode frequencies near $0.825$ and $1.069~\mathrm{MHz}$. This theoretical behavior is consistent with the measured optical output spectra, where the normal-mode splitting is unresolved at lower drive powers and develops into a clearly resolved doublet as the coupling is increased~\cite{groblacher2009observation}. The comparison therefore provides an experimental benchmark for the predicted normal-mode frequencies and spectral splitting, while a direct comparison of the relative peak heights and detailed line shapes would require an explicit model of the output homodyne detection.

The spectral quantities considered in this work are defined in terms of the intracavity optical and mechanical fluctuation operators and therefore are not all directly measured as intracavity observables. The intracavity optical field is related to the field leaving a given cavity port through the standard input-output relation, $a_{\mathrm{out},j}=a_{\mathrm{in},j} -\sqrt{\kappa_j}\,a$, where $\kappa_j$ denotes the coupling rate through that port~\cite{aspelmeyer2014cavity}. The output optical field can be analyzed using homodyne or heterodyne detection. The mechanical mode $b$, by contrast, is not detected directly. Instead, its motion is accessed through the displacement $x=x_{\mathrm{zpf}}(b+b^\dagger)$, which modulates the phase and amplitude of the cavity output field. Consequently, the experimentally accessible counterpart of the mechanical spectrum $S_{bb}(\nu)$ is the converted displacement power spectral density $S_{xx}(\nu)$ obtained through optical detection.

The optical-mechanical cross-spectra $S_{ab}(\nu)$ and $S_{ba}(\nu)$ likewise do not correspond to directly measured intracavity quantities. Experimentally, their counterparts can be obtained from cross-spectral measurements between the detected
optical-field fluctuations and the mechanically induced displacement
signal. In practice, the optical quadratures and the displacement-related signal can be recorded simultaneously, and their cross-spectral density can then be evaluated in the frequency domain. Measurements of correlations between radiation-pressure fluctuations and mechanical
motion have already been demonstrated in cavity-optomechanical systems~\cite{purdy2013observation,purdy2017quantum}. Such measurements provide an experimentally established route for probing the optical-mechanical correlations represented theoretically here by $S_{ab}(\nu)$ and $S_{ba}(\nu)$.

Experimentally, the EP can be identified by continuously tuning the drive power and tracking both the resonance frequencies and linewidths of the hybridized modes extracted from the measured optical spectra or optically detected mechanical spectra. Below the EP, the two poles share a common resonance frequency but have different decay rates, at the EP the poles coalesce, and above the EP their resonance frequencies split while their decay rates become equal. The EP is therefore determined from the simultaneous evolution of the mode frequencies and linewidths, rather than from the appearance of a resolved spectral doublet alone. Since the linearized optomechanical coupling scales as $g \propto \sqrt{n_{cav}}$, and using the parameter from above, the estimated threshold under red-sideband operation ($\Delta \simeq \omega_m$) is approximately $4.7~\mathrm{mW}$. 

Finally, the second-order correlations considered here do not require an independent experimental reconstruction within the linear Gaussian regime assumed in this work. As derived above, the absence of anomalous correlations allows the first- and second-order coherence functions to be related by the Siegert relation which applies to Gaussian field fluctuations~\cite{lassegues2022field}. Consequently, once the corresponding first-order coherence is experimentally determined, the second-order correlation predicted by the present model can be inferred directly through the Siegert relation. Alternatively, direct intensity-correlation measurements can be performed using photon-counting techniques. Optical detection combined with single-photon detection has been used to measure phonon intensity correlations and photon-phonon correlations in cavity-optomechanical systems~\cite{cohen2015phonon,riedinger2016non}.

\section{Conclusions and remarks}
\label{sec:conclusion}

We have analyzed how a second-order exceptional point (EP) governs the spectral and dynamical response of a dissipative cavity optomechanical system. At optical-mechanical resonance, $\Delta=\omega_m$, the effective non-Hermitian first-moment dynamics exhibits an EP at $\gEP=|\kappa-\gamma|/4$, where the eigenvalue splitting follows the characteristic square-root dependence on perturbations. Extending the analysis to experimentally relevant observables, we derived closed-form expressions for the steady-state moments, transient second-order moments, first-order coherence functions, spectral responses, and second-order intensity correlations. These dynamical quantities are governed by the same square-root quantity,
\[
    \Omega=\sqrt{(\kappa-\gamma)^2-16g^2}.
\]
Below the EP, the transient dynamics is non-oscillatory, whereas above it damped oscillations emerge. At the EP, the Jordan-block dynamics produces polynomial-exponential terms proportional to $e^{-\Gamma t}$, $te^{-\Gamma t}$, and $t^2e^{-\Gamma t}$ in the second-order moments.

The same EP behavior is inherited by the two-time correlations. At the EP, the first-order coherence takes the form $(A+B\tau)e^{-(\Gamma/2+\ii\omega)\tau}$, and its Fourier transform develops a second-order spectral pole. Above the EP, the spectral poles split in frequency while sharing the same decay rate, with a resolved normal-mode doublet appearing when the splitting becomes sufficiently large compared with the linewidth. Gaussian moment factorization further yields the Siegert relation, so that the zero-delay optical and mechanical autocorrelations remain fixed at the thermal value $g^{(2)}_{aa}(0)=g^{(2)}_{bb}(0)=2$, while their finite-delay behavior changes qualitatively across the EP. Overall, the population, coherence, spectral, and intensity-fluctuation signatures can be traced to the same EP physics, directly connecting mode coalescence and polynomial relaxation with experimentally accessible correlation observables.

\section*{code availability}
Python codes to generate analytical and numerical results presented in this work are available at \url{https://github.com/Khazali99/EP-OptMech-Dyn}.

\par\vspace{1.0\baselineskip}

\begin{acknowledgments}
We are grateful to the National Research and Innovation Agency (BRIN), Universitas Indonesia, and the Ministry of Higher Education, Science, and Technology of Indonesia for their financial support through the ``Degree-by-Research Scholarship'' and ``Hibah Penelitian Fundamental Reguler'' under Contract No.~PKS-176/UN2.RST/HKP.05.00/2026. We also acknowledge QuasiLab and Mahameru BRIN for providing access to mini-cluster and high-performance computing facilities.
\end{acknowledgments}

\bibliography{refs} 
\end{document}